\documentclass[dvipsnames]{faircoral}

\title{Coral Discovery: The Division of Cognitive Labor in Autonomous Research}

\author[1,*]{Xinxing Ren}
\author[2,*]{Qianbo Zang}
\author[3,*]{Ziyan Wang}
\author[1,*]{Suman Deb}
\author[4]{Qiang Yang}
\author[1]{Caelum Forder}
\author[1]{Peter Carroll}
\author[1]{Gilead Gevero}
\author[1]{Jing Liu}
\author[5]{Yali Du}
\author[3]{Milind Tambe}
\author[6,\dagger]{Zekun Guo}
\affiliation[1]{Coral AI Labs}
\affiliation[2]{Università di Trento}
\affiliation[3]{Harvard University}
\affiliation[4]{University of Cambridge}
\affiliation[5]{King's College London}
\affiliation[6]{University of Hull}
\contribution[*]{Equal contribution}
\contribution[\dagger]{Corresponding author}
\correspondence{Zekun Guo}

\abstract{
A research community is only as strong as the way it divides its cognitive labour. The philosopher Philip Kitcher argued that researchers should not all pursue the most promising approach. Credit rules, not a central plan, keep some of them on other lines. We bring this question to autonomous research, where Large Language Model agents improve programs against a deterministic grader. Agents that share only a score follow the leader and converge onto one approach. \textbf{\coral} gives the community a \textbf{direction board}. A verifier labels every attempt with a \textbf{direction}, the mechanism earning the score, and a \textbf{method}, one change within it. Per direction, the board tracks how much each eval improves the score, adds an exploration bonus for little-tried directions, prices the option of opening a new direction, and suggests how many agents each direction deserves. Credit is earned against a direction's own best, so a lagging direction keeps its agents. Agents finish their lines, then treat the board as advice. We test on nine tasks with the same model and budget, without grader access except for Kernel. \coral{} beats the strongest open autonomous multi-agent system on 6 tasks, ties on 1, and loses on 2. Against published numbers, it is best or tied on 4. It improves the reported Kernel result from 1103 to 1095 cycles and passes 1103 at eval 275, under half of the 596 used before. It sets a new Polyominoes state of the art, 91.78 against 87.0. Direction labels show that the leading scores rest on directions a score-only community never entered. Code will be released shortly.

}

\date{Oct 24, 2026}

\usepackage{hyperref}
\usepackage{url}
\usepackage{booktabs}
\usepackage{amsfonts}
\usepackage{amsmath}
\usepackage{amssymb}
\usepackage{microtype}
\usepackage{xcolor}
\usepackage{multirow}
\usepackage{array}
\usepackage{adjustbox}

\usepackage{makecell}
\usepackage{siunitx}
\usepackage{framed}
\usepackage{pifont}
\usepackage{graphicx,subcaption}
\usepackage{pifont}
\usepackage{xspace}
\usepackage{nicematrix}
\usepackage{nicefrac}
\usepackage{wrapfig}

\newcommand{\coral}{Coral Discovery}
\newcommand{\mitcoral}{MIT CORAL}
\newcommand{\dir}[1]{\texttt{#1}}

\newcommand{\logoh}{0.72cm} 
\newcommand{\cambridgeh}{0.6cm} 
\newcommand{\hullh}{0.85cm}      
\newcommand{\affillogos}{%
  \hspace{0.55cm}\raisebox{-0.5\height}{\includegraphics[height=\logoh]{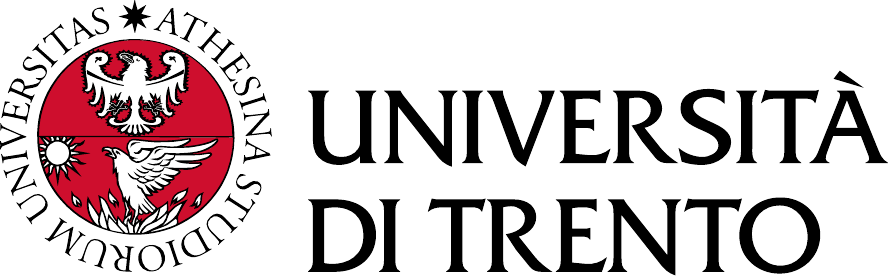}}%
  \hspace{0.45cm}\raisebox{-0.5\height}{\includegraphics[height=\logoh]{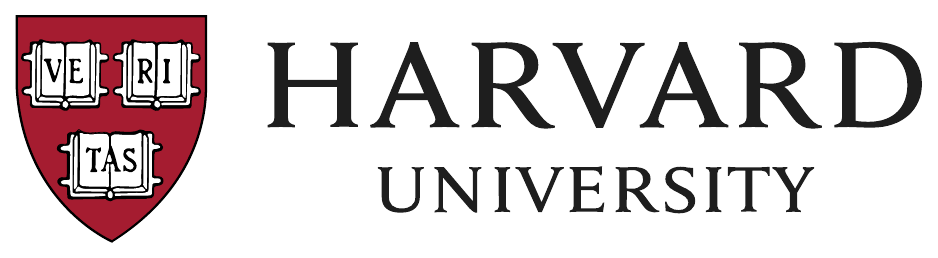}}%
  \hspace{0.45cm}\raisebox{-0.5\height}{\includegraphics[height=\cambridgeh]{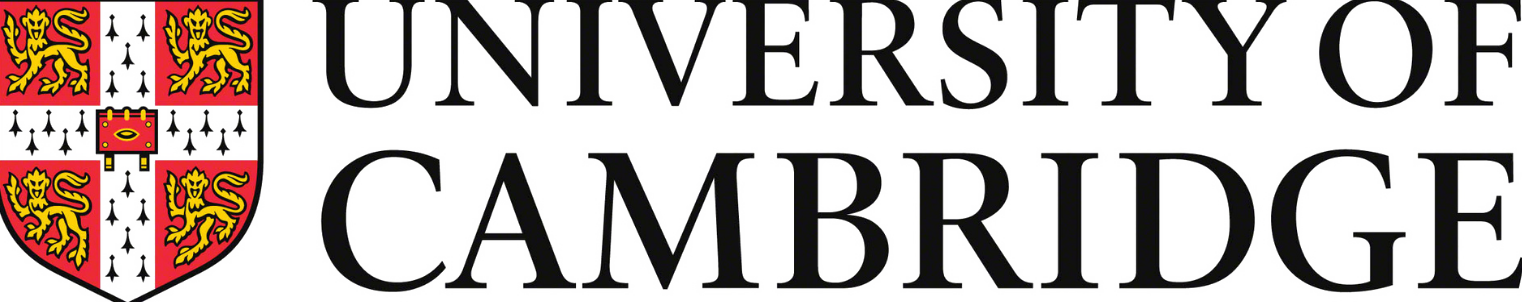}}%
  \hspace{0.45cm}\raisebox{-0.5\height}{\includegraphics[height=\logoh]{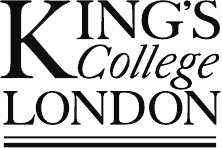}}%
  \hspace{0.45cm}\raisebox{-0.5\height}{\includegraphics[height=\hullh]{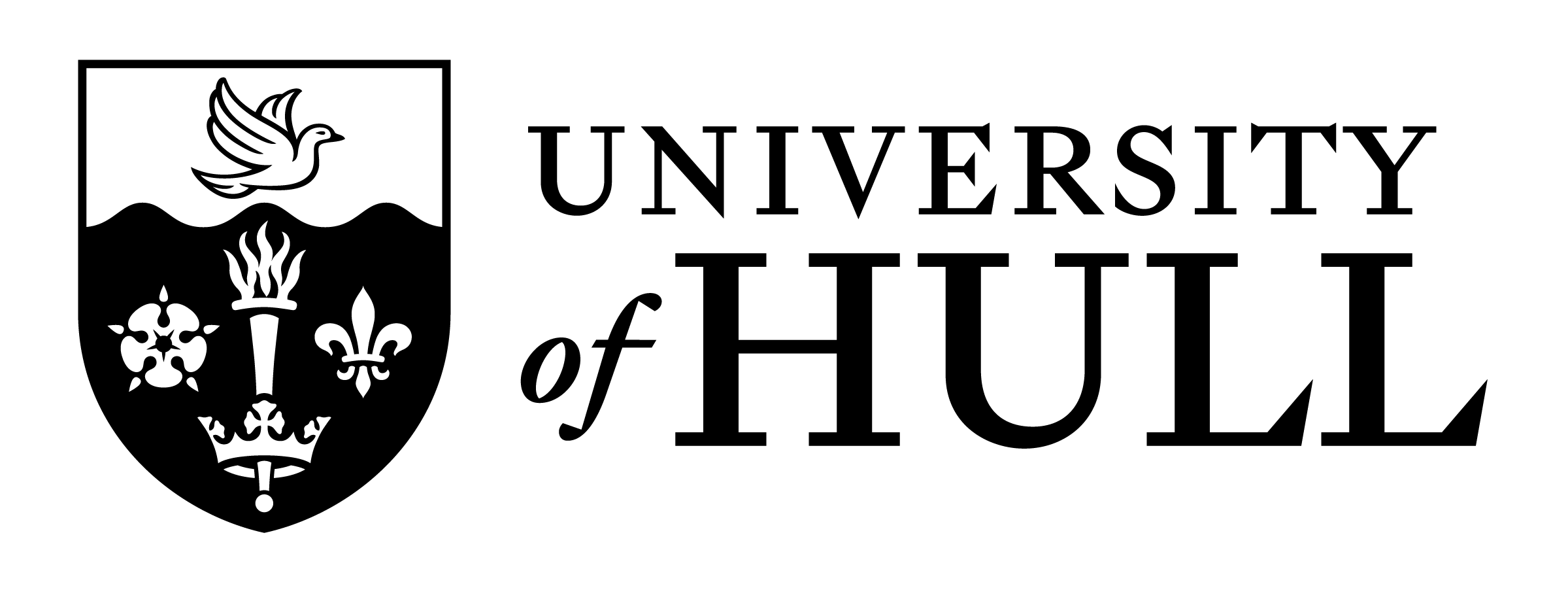}}}
\makeatletter
\patchcmd{\mymaketitle}{\includegraphics[width=0.8cm]{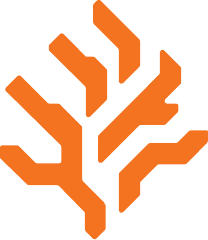}}{\raisebox{-0.5\height}{\includegraphics[width=0.8cm]{assets/coral_mark.pdf}}\affillogos}{}{\ClassError{faircoral}{logo patch failed}{}}
\makeatother

\begin{document}

\maketitle

\section{Introduction}\label{sec:intro}

\begin{figure}[t]
\centering
\includegraphics[width=\linewidth]{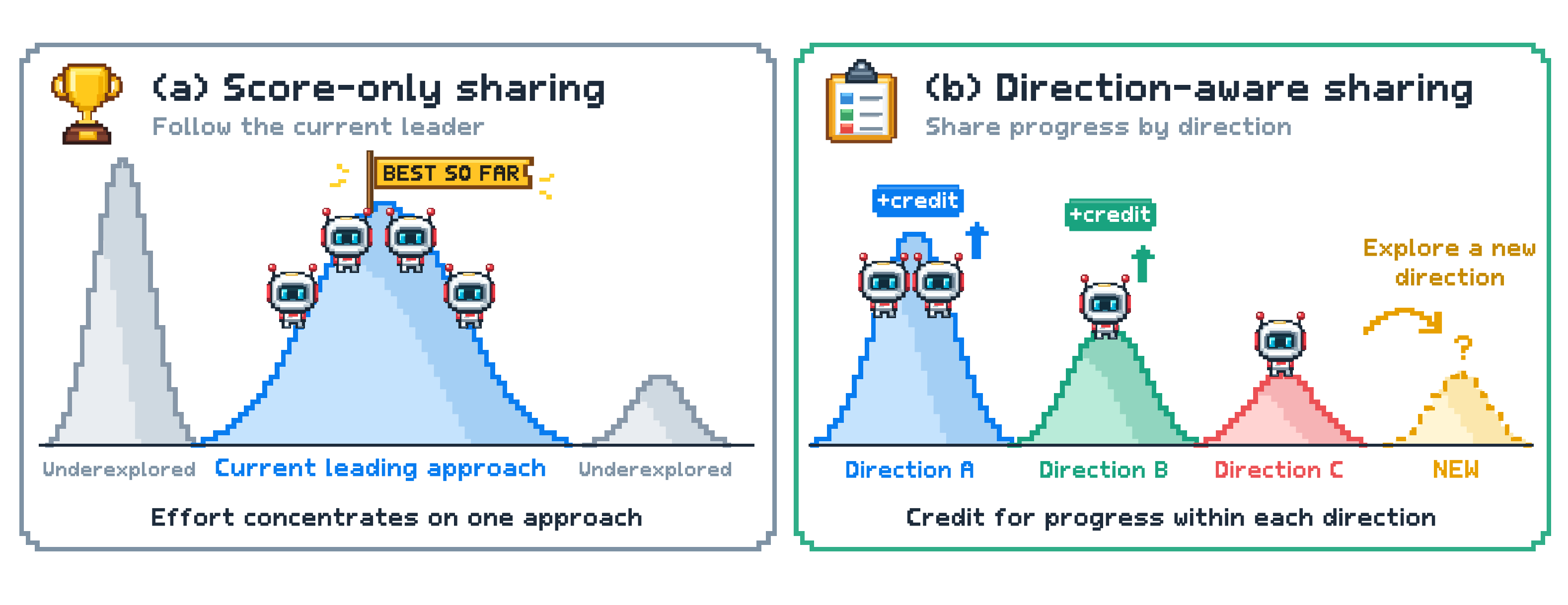}
\caption{The division of cognitive labour in a community of research agents. Left: a community that shares only a score answers ``who scores highest''. Every agent modifies the leader, and approaches that nobody has tried are never evaluated. Right: a community that shares progress per direction sees which approaches exist and how each is paying off. It spreads its agents over them, keeps opening new ones, and earns credit against each direction's own best, so a lagging direction keeps its agents. \Cref{sec:results,sec:analysis} quantify both.}
\label{fig:division}
\end{figure}

Science is done by communities, and a community's progress depends on how it divides its cognitive labour. \citet{kitcher1990division} argued that a group of researchers should not all pursue the single most promising approach. A distribution that keeps some of them on less promising lines explores more of the space. This distribution does not require a central planner. Credit rules, above all the priority that goes to whoever is first on a line, make credit-seeking researchers spread themselves over approaches. \citet{weisberg2009epistemic} made the picture concrete with an epistemic landscape. Followers who climb towards the best known result crowd onto one peak. A community that also rewards work on other peaks covers the landscape. \citet{kuhn1977essential} called the balance between pursuing the established line and striking out on a new one the essential tension of research.

Autonomous research now lets us run this question as an experiment. Many important problems have no ready-made answer but do have a definite criterion of judgement. A logistics heuristic is judged by its cost on instances \citep{chen2025heurigym}, and a kernel by its cycle count \citep{ouyang2025kernelbench}. A mathematical construction is judged by substituting it into the objective function. Placing a large language model inside an evaluator-guided search loop has become an effective way to tackle such problems. FunSearch \citep{romera2024funsearch}, AlphaEvolve \citep{novikov2025alphaevolve}, OpenEvolve \citep{sharma2025openevolve}, ShinkaEvolve \citep{lange2025shinkaevolve}, and EvoX \citep{liu2026evox} allow the model to propose candidates. A fixed evolutionary algorithm decides which parents to select, what to retain, and when to evaluate. A more recent line hands these decisions to long-running agents \citep{qu2026coral}. The agents retrieve, propose, evaluate, and write back in their own workspaces, and they collaborate through a shared memory. Several equal agents, a deterministic grader, and a shared memory form a small research community. We can observe its division of labour at every evaluation and change it by design.

Left to a shared memory alone, the community behaves like Weisberg and Muldoon's followers. The shared memory of such systems is organised around a leaderboard sorted by score. It answers ``what have others done, and who scores highest''. Facing it, an agent has two choices: to modify along the leader's line or to start from scratch on its own. In our runs, the first choice produced groups in which three out of four agents worked in the same direction for most of the run, and whole thirds of a run fell into a single direction. The second choice gave up collaboration altogether. What the leaderboard cannot answer is the question Kitcher asked. Under a limited budget, when should an agent dig deeper into an existing approach (exploitation), and when should it move to a different one (exploration)? The leaderboard has neither ``approach'' as a unit nor ``rate of return'' as a quantity.

\begin{figure}[t]
\centering
\includegraphics[width=\linewidth]{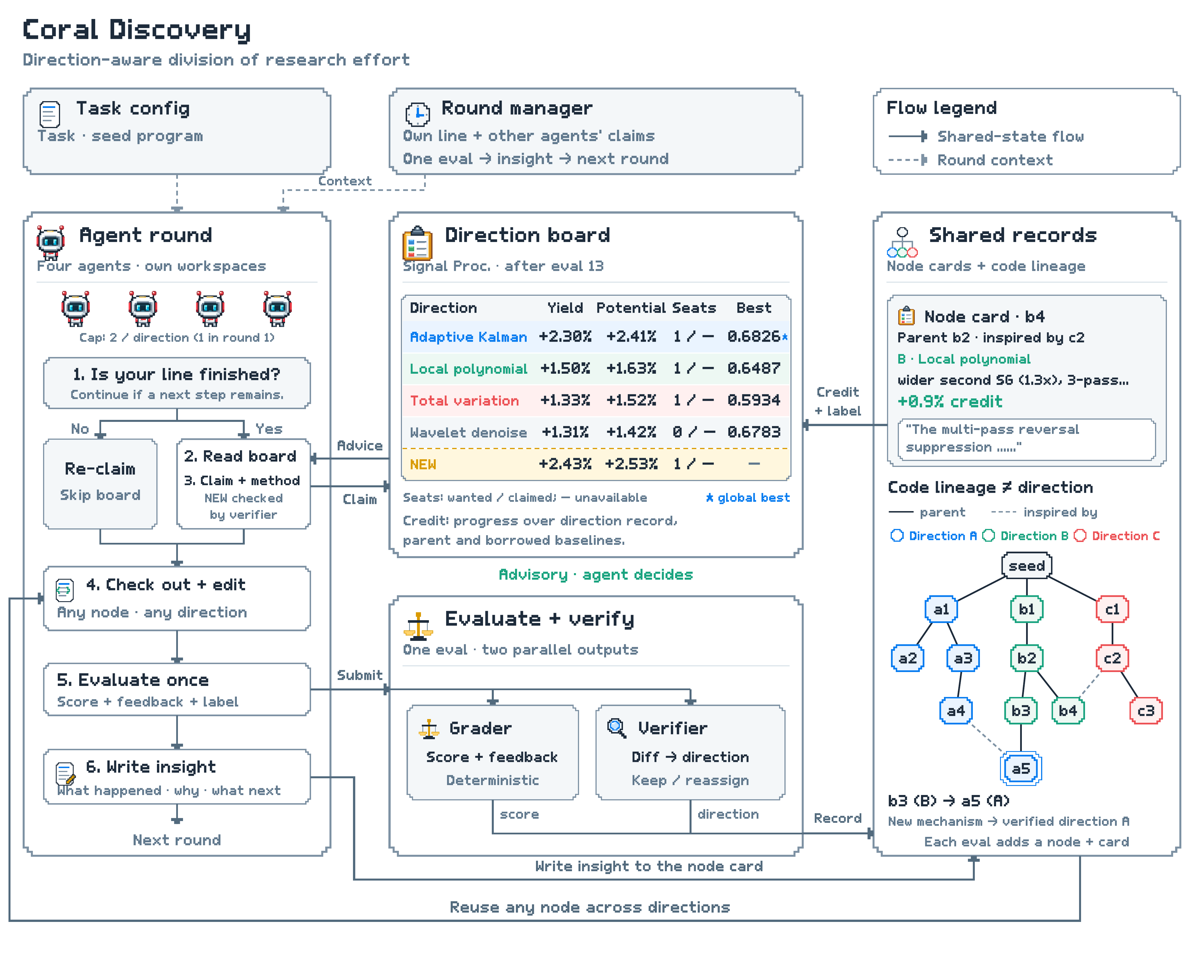}
\caption{Overview of \coral{}. Left: one round of an agent (1 is the line finished, 2 read the board, 3 claim, 4 check out and edit, 5 eval, 6 insight), an unfinished line loops back and skips the board. Middle: the shared direction board, the community's account of its division of labour, and the verifier, which files every eval by its diff. Right: the tree of node cards, where code lineage and direction labels are independent graphs, and a lagging direction still earns credit for its own records. Board numbers are from the Signal Proc. run after eval 13.}
\label{fig:overview}
\end{figure}

We propose \textbf{\coral{}}, a direction board that provides a community of research agents with this missing unit and this missing quantity. A credit rule keeps lagging approaches alive (\Cref{fig:division}). We maintain the setting of autonomous multi-agent evolution. A task is defined by a description $x$ and an evaluator $E$. Evaluating a candidate $y$ returns $E(x, y) = (s, f)$ with a score $s$ and feedback $f$. $N$ agents run asynchronously in their own workspaces, each repeating ``retrieve context $\rightarrow$ propose a candidate $\rightarrow$ evaluate $\rightarrow$ write back to shared memory''. We add three components, summarised in \Cref{fig:overview}.

\textbf{Two-level labelling into directions and methods, plus a verifier.} Each attempt carries two labels. The direction, an approach in Kitcher's sense, is the mechanism by which the program earns its score. A verifier determines it from reading the diff. The method, what was changed this time, is written by the agent itself. The direction is the unit of budget allocation and return statistics. The method is the working record inside a direction.

\textbf{The direction board.} For each direction the board maintains four quantities. Yield is how much improvement each eval spent buys. Potential is what the direction is worth once the uncertainty of having been tried little is added. Wanted is a suggested head count allocated by potential, and claimed is how many agents are on it now. It also treats opening another direction as an option, NEW, whose value is estimated from the past performance of new directions. Credit is measured against a direction's own best. This is the priority rule of the community: a record inside a lagging direction is worth as much as a record inside the leading one.

\textbf{The round protocol.} In each round, an agent first judges whether its own line is finished. It reads the board only once it is, claims a direction and a method, edits freely, runs one eval and writes an insight. The board only directs attention, and the decision rests with the agent. The verifier intervenes at two points, claim and eval.

\Cref{tab:paradigms} places the three paradigms side by side. Fixed evolutionary search hard-codes the division of labour. Autonomous multi-agent evolution leaves it to each agent alone. \coral{} gives the community a shared account of it (details for each cell are given in \Cref{sec:method}).

\paragraph{Contributions.}
\begin{itemize}
\item \textbf{A two-level abstraction of directions and methods, and a verifier.} We define a direction as the mechanism by which a program earns its score, and a method as a single change within a direction. An independent model session, the verifier, determines the direction from the program diff at every eval. The direction is the unit of budget allocation and return statistics. This makes comparison across agents and analysis across runs possible (\Cref{sec:method,sec:phenomena}).
\item \textbf{The direction board and the round protocol.} The board maintains yield, potential and wanted for each direction, and treats opening another direction as an option, NEW, whose value can be estimated. The protocol requires four things. An agent finishes its own line before reading the board. Claims pass a head-count cap and a duplicate check. Each round contains one eval, and an insight is written after each eval. The numbers on the board are advisory, not binding (\Cref{sec:method}).
\item \textbf{Strictly controlled experiments on nine tasks.} We allow no internet access and no reading of the grader (except for Kernel), under the same model and the same four-agent budget. \coral{} beats the strongest open autonomous multi-agent system, \mitcoral{} \citep{qu2026coral}, on 6 tasks, ties on 1, and falls behind on 2 (LLM-SQL, Txn Sched.). Against all numbers published for these tasks, it achieves the best or tied-best result on 4. Under the same conditions, it improves the reported Kernel result from 1103 to 1095 cycles. It passes 1103 at eval 275, fewer than half of the 596 evals used before. It also sets a new state of the art on Polyominoes, 91.78 against the previous 87.0 (\Cref{sec:results}).
\item \textbf{Attribution analysis.} We label every submission of every run with a direction. The scores by which \coral{} leads come from directions a score-only community never entered. These directions were opened and kept alive until they produced a record. We attribute this, respectively, to the claim rules and the NEW arm, to per-direction credit, and to the inspired-by threshold. We also report the cost of exploration faithfully (\Cref{sec:analysis,sec:limitations}).
\end{itemize}

\begin{table}[t]
\caption{Three paradigms of evaluator-guided search and how each divides the community's labour. Fixed evolutionary search covers FunSearch \citep{romera2024funsearch}, AlphaEvolve \citep{novikov2025alphaevolve}, OpenEvolve \citep{sharma2025openevolve}, ShinkaEvolve \citep{lange2025shinkaevolve} and EvoX \citep{liu2026evox}. Autonomous multi-agent evolution follows \citet{qu2026coral}.}
\label{tab:paradigms}
\centering
\scriptsize
\begin{adjustbox}{max width=\linewidth}
\begin{tabular}{@{}>{\raggedright\arraybackslash}p{0.15\linewidth}>{\raggedright\arraybackslash}p{0.24\linewidth}>{\raggedright\arraybackslash}p{0.24\linewidth}>{\raggedright\arraybackslash}p{0.31\linewidth}@{}}
\toprule
 & Fixed evolutionary search & Autonomous multi-agent evolution & \coral{} (this work) \\
\midrule
Who decides what to look at & Fixed rules select parents and assemble the context & The agent reads attempts / notes / skills itself & The agent reads the board, directions and nodes, aggregated by direction \\
Who decides what to do & The LLM generates one mutation given the context & The agent plans, implements and tests & The agent claims a direction and a method, then implements. Round 1: one direction per agent \\
Mode of collaboration & Stateless parallel evaluation & Indirect collaboration through shared memory & Through the tree: every node carries a direction, a method and an insight. Building on others' nodes is allowed and declared \\
Exploration / exploitation & Hard-coded in the algorithm (MAP-Elites, islands) & Each agent judges alone, the heartbeat prompts a change on a stall & Return estimates per direction and for NEW (advisory). At most two agents per direction. Diagnose on a stall. The agent decides \\
\bottomrule
\end{tabular}
\end{adjustbox}
\end{table}

\section{Related Work}\label{sec:related}

\paragraph{Automated research with LLMs.}
FunSearch \citep{romera2024funsearch} and AlphaEvolve \citep{novikov2025alphaevolve} place a language model inside an evaluator-guided evolutionary loop. OpenEvolve \citep{sharma2025openevolve}, ShinkaEvolve \citep{lange2025shinkaevolve}, EvoX \citep{liu2026evox}, AdaEvolve \citep{cemri2026adaevolve}, GEPA \citep{agrawal2025gepa} and TTT-Discover \citep{yuksekgonul2026ttt} refine that loop, but a fixed algorithm still decides what to retain. The AI Scientist \citep{lu2024aiscientist}, the AI co-scientist \citep{gottweis2025coscientist} and \mitcoral{} \citep{qu2026coral} hand the loop to autonomous agents. \coral{} keeps the autonomous setting and adds the layer that divides the community's evaluations over approaches.

\paragraph{Multi-agent systems.}
MetaGPT \citep{hong2024metagpt}, ChatDev \citep{qian2024chatdev}, AutoGen \citep{wu2024autogen} and CAMEL \citep{li2023camel} trade one long context for division of labour and negotiation. AgentRadio \citep{ren2026agentradio} shows that the gains concentrate on decomposable tasks, and that interdependent subtasks need coordination during execution. \coral{} divides a single objective over directions instead of subtasks. A shared, verified ledger replaces negotiation.

\paragraph{Exploration and exploitation.}
UCB1 \citep{auer2002ucb}, UCT \citep{kocsis2006uct,coulom2006mcts,browne2012mcts} with progressive widening as in AlphaGo \citep{silver2016alphago}, and quality-diversity methods such as MAP-Elites \citep{mouret2015mapelites} are the standard tools. The board treats directions as arms. Yield is a windowed mean, potential is a UCB-style bonus on lifetime counts, and NEW is an expansion arm. Two differences matter. The reward is a direction's own new high rather than the raw score, and the policy is advisory, since the agent reads the evidence and decides.

\section{Method: Coral Discovery}\label{sec:method}

\coral{} adds directions and methods with their verifier (\Cref{sec:directions}), the direction board (\Cref{sec:board}) and the round protocol (\Cref{sec:protocol}).

\subsection{Directions and Methods}\label{sec:directions}

\paragraph{Definitions.}
A \textbf{direction} is the mechanism by which a program earns its score. It has three parts: the objective or proxy optimised, the solution object or decision unit, and the principle of search or construction. Programs with one mechanism share a direction, however they are written. A \textbf{method} is a concrete change within a direction: a parameter, variant, reordering, substitution or speed-up. The verifier assigns the direction, which is the unit of allocation and return statistics. The agent supplies the method on the node card.

\begin{figure}[t]
  \centering
  \includegraphics[width=\linewidth]{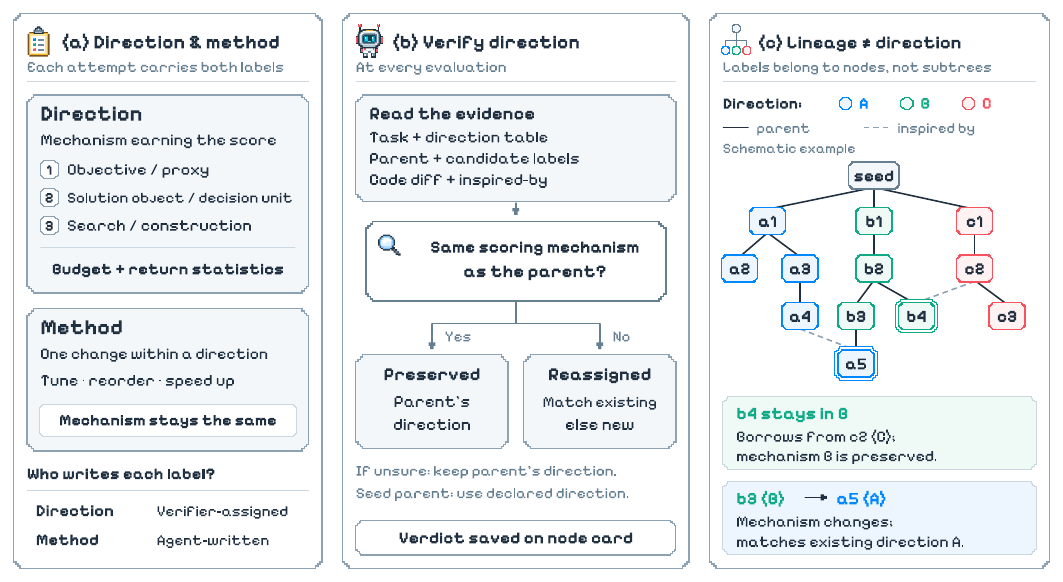}
  \caption{Directions and methods. Left: the two definitions, with a direction being an approach in Kitcher's sense. Middle: the verifier's judgement at every eval, with its verdict recorded on the node card. Right: code lineage (solid edges) and direction membership (node colours) are independent; borrowing is declared with \texttt{--inspired-by} (dashed edges), and direction labels attach to nodes rather than subtrees.}
  \label{fig:directions}
\end{figure}

\paragraph{Verifier.}
The verifier is a tool-less single-turn session of the agents' model, run at every \texttt{coral eval}. It reads the task description, the direction table, the (direction, method) pairs of parent and candidate, the diff and the declared inspired-by nodes. It answers one question: does the candidate preserve the mechanism by which the parent earns its score? If so, the candidate inherits the parent's direction. Otherwise it joins a matching or new direction, and provisional duplicates of earlier mechanisms are merged. The verdict defaults to ``preserved'' when uncertain and goes on the node card. Calls run serially, and a 5-minute timeout keeps the agent's declaration. Code lineage and direction membership are independent graphs (\Cref{fig:directions}). A direction labels nodes, not subtrees.

\subsection{The Direction Board}\label{sec:board}

The board keeps four quantities per direction. Let $W$ be the window length, $N$ the evals in the window, $n_d$ and $n_d^{\mathrm{all}}$ the evals of direction $d$ in the window and over the run, and $m$ the mean credit in the window. Failed evals have credit 0, and minimisation scores are negated. Defaults are $W=32$, $\mathrm{prior}_n=4$, $k=3$, $\kappa=0.05$ and $n_{\mathrm{assign}}$ the number of agents.

\paragraph{Credit.}
The credit of node $X$ is its score above a threshold, normalised by the current global best (\Cref{eq:credit}):
\begin{equation}
  \operatorname{credit}(X) = \frac{\max\!\Bigl(0,\, s(X) - \max\bigl(s(\mathrm{parent}),\ \max_{i \in \text{inspired-by}} s(i),\ \mathrm{best}_{d(X)}\bigr)\Bigr)}{\lvert \text{global best} \rvert}
  \label{eq:credit}
\end{equation}
The threshold includes the direction best, so only a direction record earns credit. Copying a better program merely raises the threshold. Directions are therefore compared on what they add rather than on what they borrow. Normalisation keeps the constants below task-independent.

\paragraph{Yield.}
The yield of direction $d$ is its mean credit per eval in the window, shrunk towards the overall mean with $\mathrm{prior}_n$ pseudo-evals (\Cref{eq:yield}):
\begin{equation}
  \mathrm{yield}_d = \frac{\sum_{X \in \text{window},\, d(X) = d} \operatorname{credit}(X) + \mathrm{prior}_n \cdot m}{n_d + \mathrm{prior}_n}
  \label{eq:yield}
\end{equation}
Shrinkage keeps a sparsely sampled direction near the mean. The window returns a stalled direction there within $W$ evals.

\paragraph{Potential.}
Potential adds an exploration bonus to yield (\Cref{eq:potential}):
\begin{equation}
  \mathrm{potential}_d = \mathrm{yield}_d + c \sqrt{\ln N / (n_d^{\mathrm{all}} + 1)}, \qquad c = \kappa \, y^{+}
  \label{eq:potential}
\end{equation}
where $y^{+}$ is the highest yield in the run. Whole-run counts keep an abandoned direction from looking untried. Because $y^{+}$ never decays, the bonus keeps proposing probes after all directions stall.

\paragraph{NEW.}
How many directions a task rewards is unknown in advance. The NEW arm estimates it from the run. If new directions gained soon after opening, opening more pays. Otherwise depth pays. Consider every direction $d$ opened after a global best $G_d$ existed. Let $g_d$ be the improvement over $G_d$ of its best score in its first $k$ evals (its infancy), truncated at 0, and let $n_d^{0} = \min(k, n_d^{\mathrm{all}})$. \Cref{eq:new} gives the estimate and its potential:
\begin{equation}
  \hat{y}_{\mathrm{NEW}} = \frac{\sum_d g_d + k \, y^{+}}{\sum_d n_d^{0} + k}, \qquad
  \mathrm{potential}_{\mathrm{NEW}} = \hat{y}_{\mathrm{NEW}} + c \sqrt{\ln N / \bigl(\textstyle\sum_d n_d^{0} + 1\bigr)}
  \label{eq:new}
\end{equation}
With $k$ pseudo-evals of prior $y^{+}$, $\hat{y}_{\mathrm{NEW}}$ equals $y^{+}$ before any opening, and $y^{+}/(j+1)$ after $j$ openings without an infancy record. NEW therefore starts high and decays as the global best rises.

\paragraph{Wanted.}
Each round, $n_{\mathrm{assign}}$ seats go to the directions and NEW in proportion to potential (largest remainder). When every potential is 0, all seats go to the global-best direction. A direction whose wanted exceeds its claimed is under-served.

\paragraph{Advisory, not binding.}
Wanted is not enforced, because the four quantities are statistics. A high-yield direction may have room or be exhausted. A long-untouched direction lifted by the bonus may be dead or promising. Statistics cannot tell these apart. They point to the directions worth inspecting, and the judgement after reading attempts, insights and diffs is the agent's.

\subsection{Protocol}\label{sec:protocol}

Agents touch the shared state only through \texttt{coral} commands, since native \texttt{git} is disabled in their workspaces. We inherit the command set of \mitcoral{} (\texttt{coral checkout}, \texttt{coral show}, \texttt{coral diff}, \texttt{coral skills} and \texttt{coral eval}). We add five commands for the board and the tree (\texttt{coral board}, \texttt{coral direction}, \texttt{coral claim}, \texttt{coral insight}, \texttt{coral tree show}). We extend \texttt{coral eval} with a declared direction and inspired-by nodes, and we withdraw the leaderboard commands. \Cref{sec:commands} lists the complete set. Each step below names the command it uses.

The manager drives each round in two phases. It injects only the agent's \textbf{own line} (direction, method, last 5 scores and bests) and \textbf{the other agents' current claims}. (1) After a line best, or with a concrete next step, the agent re-claims its direction with \texttt{coral claim} without reading the board. After 5 scored evals without a personal best, a Diagnose block asks whether the bottleneck is structural or removable. (2) Otherwise it reads the board with \texttt{coral board}, a direction's attempts with \texttt{coral direction \textless{}id\textgreater{}}, a node card with \texttt{coral tree show \textless{}commit\textgreater{}} and a diff with \texttt{coral show \textless{}hash\textgreater{} --diff}. (3) \texttt{coral claim \textless{}id\textgreater{} --method "\ldots"} claims a direction. \texttt{coral claim --new "\textless{}name\textgreater{}" --description "\ldots" --method "\ldots"} opens a provisional one. Both pass three checks: no method repeats another agent's verbatim, at most 2 agents per direction (1 in the first round), and no \texttt{--new} duplicating an existing mechanism. (4) \texttt{coral checkout \textless{}hash\textgreater{}} takes any node of any direction as parent, and the agent edits freely. Nodes it borrowed from are declared at eval time with \texttt{--inspired-by}, which enters them into the credit threshold. (5) \texttt{coral eval -m "\ldots" --direction \textless{}id\textgreater{} [--inspired-by c1,c2]} commits, writes the node card and releases the claim. Scoring and the verifier run in parallel. A round has one eval. Once both results arrive, the manager interrupts the session and starts phase 2. (6) \texttt{coral insight \textless{}commit\textgreater{} "\ldots"} records what happened, why and what it implies. It is the only write operation of phase 2.

\section{Experimental Setup}\label{sec:setup}

\subsection{Tasks, Baselines and Cited Numbers}\label{sec:tasks}

Tasks and scoring follow the \mitcoral{} paper \citep{qu2026coral}. The nine tasks fall into five groups. Kernel optimisation: Kernel Eng., a VLIW SIMD kernel from \citet{anthropic2025kernel} scored in cycles. Combinatorial packing: Polyominoes \citep{pmlr-v306-mang26a} and Circle-Pack., both maximising a packing objective. Analysis and combinatorics: Signal Proc., a filter scored on held-out signals, and Erd\H{o}s Over., the minimum overlap constant. LLM inference systems: EPLB, expert load balancing, and LLM-SQL, query ordering for prefix caching. Data and cloud systems: Txn Sched., transaction scheduling for makespan, and Cloudcast, multi-cloud broadcast cost. The baselines are \mitcoral{} under our conditions: one Claude Code agent, or four sharing a repository and a leaderboard. \coral{} adds the direction board (\Cref{sec:board}) to the same four agents and eval budget. We also cite numbers for SOTA, OpenEvolve (OE) \citep{sharma2025openevolve}, ShinkaEvolve (SE) \citep{lange2025shinkaevolve}, EvoX \citep{liu2026evox} and CORAL (single-agent, with 4-agent numbers only on Polyominoes and Kernel). They are marked \textsuperscript{\dag} in \Cref{sec:table}.

\subsection{Experimental Conditions}\label{sec:conditions}

The \mitcoral{} repository by default allows \textbf{web research} and \textbf{reading the grader}. Some graders, however, contain data an agent should not see. The Signal Proc.\ grader holds the scored test signals, and the systems graders hold the workloads on which a policy is scored. All three methods therefore run with \textbf{no internet access and no reading of the grader}. \texttt{WebSearch} and \texttt{WebFetch} are disabled, the grader directory is on the \texttt{Read} deny list, and the agent prompt states both rules. The only exception is Kernel Eng., whose grader is the VLIW simulator the task is about. We scanned the tool calls of every run afterwards and found no attempt by any agent to access the internet or to read a grader. The model is Claude Opus 4.6 (effort medium).

\subsection{Evaluation and Reporting}\label{sec:reporting}

Each configuration ran 3 times. \Cref{sec:table} reports the best run (best of 3, the counterpart of \mitcoral{}'s Bo4) with its eval count (crashes and timeouts included) and \texttt{best@k}, the first eval reaching the run's final best. Except for kernel, only the first 100 evals count. Cloudcast counts the first 20, within which all three methods reach the optimal cost 618. NA marks a missing run.

\section{Results}\label{sec:results}

\subsection{Main results}\label{sec:table}

\begin{figure}[t]
\centering
\includegraphics[width=\linewidth]{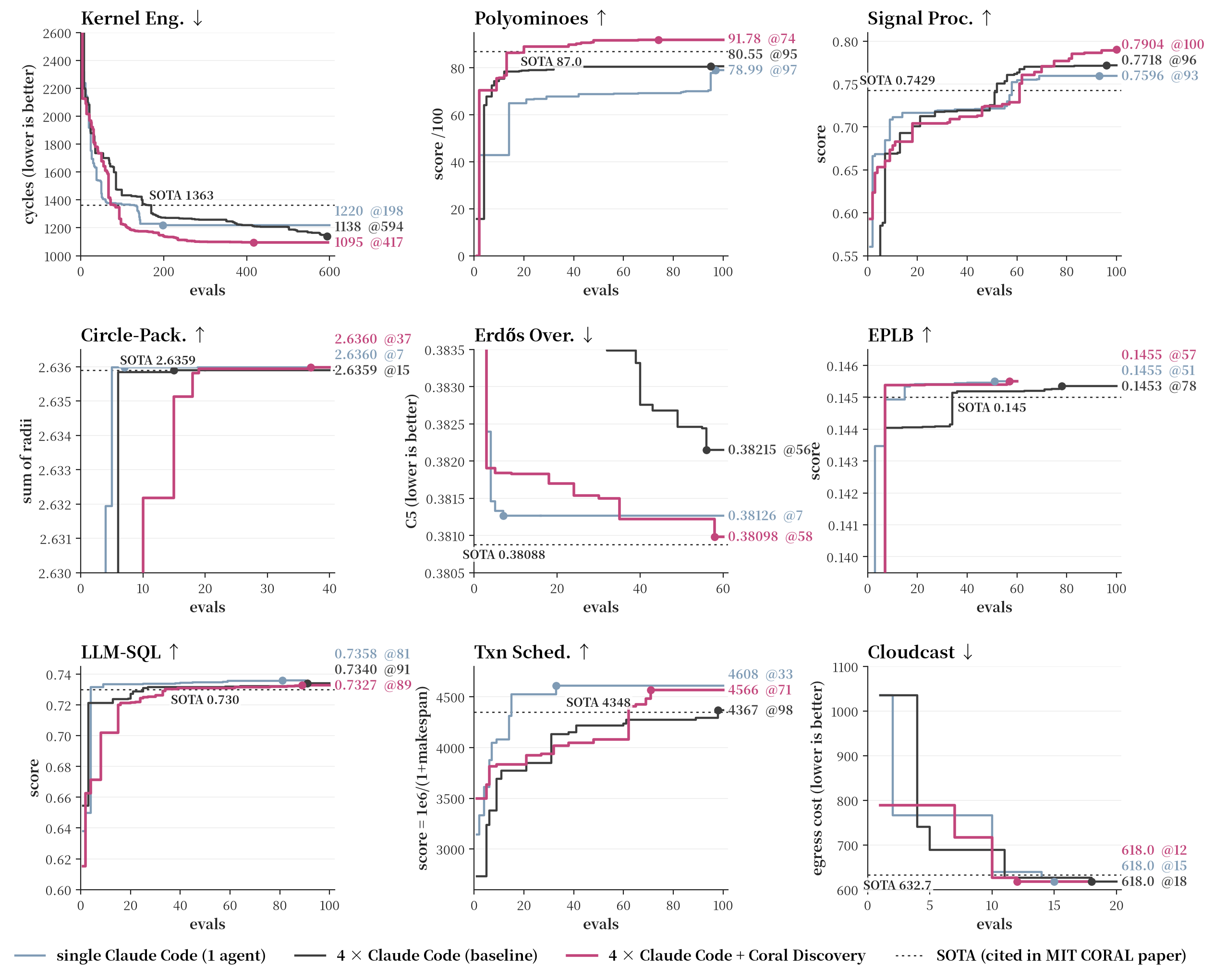}
\caption{Best-so-far curves of the single agent, the 4-agent baseline and \coral{} on the nine tasks (arrows give the direction of improvement). Dots mark the eval at which each final best first appears, labels give the final value and \texttt{best@k}, and the dashed line is the SOTA cited in the \mitcoral{} paper. Cloudcast shows the first 20 evals, within which all three methods reach 618.}
\label{fig:trajectories}
\end{figure}

\Cref{tab:main} gathers the nine tasks. Six columns are cited from the \mitcoral{} paper \citep{qu2026coral}, and three come from our Opus~4.6 runs under the conditions of \Cref{sec:conditions}. \Cref{fig:trajectories} shows the corresponding best-so-far curves. They are uncapped for Kernel Eng., cover the first 20 evals for Cloudcast, and the first 100 evals otherwise.

\begin{table}[t]
\providecommand{\ourbest}[1]{\textcolor{corallink}{\textit{#1}}}
\centering
\caption{Main results on the nine \mitcoral{} tasks. The six left-hand columns are cited from the \mitcoral{} paper\textsuperscript{\textdagger} and the three right-hand columns are our own runs with Opus~4.6\textsuperscript{\textdaggerdbl}. Arrows give the direction of improvement, \textbf{bold} marks the best across all columns of a task, and \ourbest{orange italic} marks the best among our three methods.}
\label{tab:main}
\scriptsize
\setlength{\tabcolsep}{3pt}
\begin{adjustbox}{max width=\linewidth}
\begin{tabular}{ll*{9}{c}}
\toprule
 & & \multicolumn{6}{c}{Cited from the \mitcoral{} paper\textsuperscript{\textdagger}} & \multicolumn{3}{c}{Our runs (Opus~4.6, no internet, no grader access)\textsuperscript{\textdaggerdbl}} \\
\cmidrule(lr){3-8} \cmidrule(lr){9-11}
 & Task & SOTA & OE & SE & EvoX & \makecell{CORAL \\ (single agent)} & \makecell{CORAL \\ (4 agents)} & \makecell{CORAL \\ (single agent)} & \makecell{CORAL \\ (4 agents)} & \makecell{4 $\times$ Claude Code \\ + \coral} \\
\midrule
Kernel optimisation & Kernel Eng.\ $\downarrow$ & 1363 & 2740 & NA & NA & 1350 & 1103 & \makecell{1220 \\ {\tiny 600 evals, best@198}} & \makecell{1138 \\ {\tiny 600 evals, best@594}} & \makecell{\textbf{\ourbest{1095}} \\ {\tiny 596 evals, best@417}} \\
\midrule
\multirow{2}{*}{\makecell[l]{Combinatorial \\ packing}} & Polyominoes $\uparrow$ & 87.0 & NA & NA & NA & 80.2 & 84.2 & \makecell{78.99 \\ {\tiny 100 evals, best@97}} & \makecell{80.55 \\ {\tiny 100 evals, best@95}} & \makecell{\textbf{\ourbest{91.78}} \\ {\tiny 100 evals, best@74}} \\
 & Circle-Pack.\ $\uparrow$ & 2.6359 & 2.6293 & 2.6001 & 2.6320 & \textbf{2.6360} & NA & \makecell{2.635977 \\ {\tiny 40 evals, best@7}} & \makecell{2.635890 \\ {\tiny 40 evals, best@15}} & \makecell{\textbf{\ourbest{2.635983}} \\ {\tiny 40 evals, best@37}} \\
\midrule
\multirow{2}{*}{\makecell[l]{Analysis and \\ combinatorics}} & Signal Proc.\ $\uparrow$ & 0.7429 & 0.6420 & 0.8171 & 0.7306 & \textbf{0.8229} & NA & \makecell{0.7596 \\ {\tiny 100 evals, best@93}} & \makecell{0.7718 \\ {\tiny 100 evals, best@96}} & \makecell{\ourbest{0.7904} \\ {\tiny 100 evals, best@100}} \\
 & Erd\H{o}s Over.\ $\downarrow$ & \textbf{0.38088} & 0.38188 & 0.38156 & 0.38125 & 0.38089 & NA & \makecell{0.381264 \\ {\tiny 60 evals, best@7}} & \makecell{0.382150 \\ {\tiny 60 evals, best@56}} & \makecell{\ourbest{0.380984} \\ {\tiny 60 evals, best@58}} \\
\midrule
\multirow{2}{*}{\makecell[l]{LLM inference \\ systems}} & EPLB $\uparrow$ & 0.145 & 0.127 & 0.129 & 0.146 & \textbf{0.149} & NA & \makecell{0.145498 \\ {\tiny 60 evals, best@51}} & \makecell{0.145346 \\ {\tiny 100 evals, best@78}} & \makecell{\ourbest{0.145501} \\ {\tiny 60 evals, best@57}} \\
 & LLM-SQL $\uparrow$ & 0.730 & 0.716 & 0.724 & 0.726 & 0.731 & NA & \makecell{\textbf{\ourbest{0.735769}} \\ {\tiny 90 evals, best@81}} & \makecell{0.733963 \\ {\tiny 100 evals, best@91}} & \makecell{0.732704 \\ {\tiny 100 evals, best@89}} \\
\midrule
\multirow{2}{*}{\makecell[l]{Data and cloud \\ systems}} & Txn Sched.\ $\uparrow$ & 4348 & 3774 & 3802 & 3984 & 4566 & NA & \makecell{\textbf{\ourbest{4608}} \\ {\tiny 100 evals, best@33}} & \makecell{4367 \\ {\tiny 100 evals, best@98}} & \makecell{4566 \\ {\tiny 100 evals, best@71}} \\
 & Cloudcast $\downarrow$ & 632.7 & 627.2 & 627.8 & 623.5 & 618.4 & NA & \makecell{\textbf{\ourbest{618.0}} \\ {\tiny 18 evals, best@15}} & \makecell{\textbf{\ourbest{618.0}} \\ {\tiny 20 evals, best@18}} & \makecell{\textbf{\ourbest{618.0}} \\ {\tiny 18 evals, best@12}} \\
\bottomrule
\end{tabular}
\end{adjustbox}
\par\vspace{3pt}
\begin{minipage}{\linewidth}\scriptsize\raggedright
\textsuperscript{\textdagger} From the \mitcoral{} paper (OE = OpenEvolve, SE = ShinkaEvolve), whose runs allowed web access and grader reading. Its CORAL column is the single-agent run, with 4-agent numbers only for Polyominoes and Kernel. Best Known 1363 is the Anthropic value cited there (the public leaderboard has since moved with human-in-the-loop entries and is not comparable). \textsuperscript{\textdaggerdbl} Our runs under the conditions of \Cref{sec:conditions}, evals and best index on the second line. \textbf{Bold} marks the best of all columns at the cited precision (among our runs only the best at full precision), \ourbest{orange italic} the best of our three runs. Lower is better for Kernel, Erd\H{o}s and Cloudcast.
\end{minipage}
\end{table}

\paragraph{Against the reproduced \mitcoral{} system.}
Under identical conditions, model and budget, \coral{} scores \textbf{6 wins} (Kernel Eng., Polyominoes, Signal Proc., Circle-Pack., Erd\H{o}s Over., EPLB), \textbf{1 tie} (Cloudcast) and \textbf{2 losses} (LLM-SQL, Txn Sched.) against the better of our single agent and 4-agent baseline. Four tasks show a clear margin: Kernel 1095 vs.\ 1138, Polyominoes 91.78 vs.\ 80.55, Signal 0.7904 vs.\ 0.7718, and Erd\H{o}s 0.380984 vs.\ 0.381264. On Circle-Pack.\ and EPLB the margin lies beyond the third decimal place. On Txn Sched.\ \coral{} beats the 4-agent baseline but not the single agent. On LLM-SQL it trails both (\Cref{sec:phenomena,sec:analysis,sec:limitations}). Adding agents alone brings no improvement. The 4-agent baseline beats the single agent on only three tasks, so the gain comes from the direction board, not from the agent count. The lead also arrives earlier. On Kernel Eng.\ \coral{} reached the baseline's final best (1138 cycles) at eval 202, whereas the baseline reached it at eval 594. On the other tasks it reached the baseline's final best between eval 3 (Erd\H{o}s Over.) and eval 75 (Signal Proc.). The exception is LLM-SQL, where it never did, and the baseline instead reached Discovery's final best at eval 91. On no other task did the baseline reach Discovery's final best.

\paragraph{Against published numbers.}
Across all columns of \Cref{tab:main} (bold cells), \coral{} is best or tied on \textbf{4 tasks} (Kernel Eng., Polyominoes, Circle-Pack., Cloudcast), although it ran without web access or grader reading (kernel excepted). It is sole best on Kernel Eng.\ and Polyominoes. Polyominoes 91.78 exceeds the cited SOTA 87.0. On Kernel, with the same model and four-agent budget, \coral{} improves the \mitcoral{} result from 1103 to 1095 cycles. It passes 1103 at eval 275, under half of the paper's 596 evals. Head to head against the paper's CORAL column, \coral{} has 4 wins, 2 ties and 3 losses (Signal Proc., Erd\H{o}s Over., EPLB). All three losses are to numbers obtained with web access and grader reading. The Signal grader contains the test signals, so the paper's 0.8229 is not comparable with our 0.7904, which exceeds the cited SOTA 0.7429. The Erd\H{o}s and EPLB gaps lie in the fourth and third decimal place.


\subsection{Where the lead comes from}\label{sec:phenomena}

Not all three methods ran a direction board. To compare them, we re-labelled every submission with the verifier's criterion (Opus 4.6 reading each program diff) and aligned the resulting directions per task. \Cref{tab:phenomena} reports three things: how many directions Discovery ran that the 4-agent baseline never evaluated, Discovery's margin over the baseline, and how much of this gain comes from Discovery-only directions.

\begin{table}[t]
\centering
\caption{Where Discovery's gain over the 4-agent baseline falls, by re-labelled direction: number of Discovery-only directions (those the baseline never ran) and how many produced records, the gain over the baseline in task units ($\downarrow$ tasks sign-flipped), the share of that gain from records in Discovery-only directions (own) and from records built on them (lineage), and Discovery-only over all directions in the winning lineage.}
\label{tab:phenomena}
\small
\setlength{\tabcolsep}{3pt}
\begin{adjustbox}{max width=\linewidth}
\begin{tabular}{@{}lccccc@{}}
\toprule
Task & \makecell{Discovery-only \\ directions} & \makecell{Discovery $-$ \\ baseline} & \makecell{Gain share, \\ own records} & \makecell{Gain share, \\ lineage} & \makecell{Discovery-only / all \\ lineage directions} \\
\midrule
Kernel Eng.\ $\downarrow$ & 25 (7 produced records) & \makecell{+43 \\ {\scriptsize (43 fewer cycles)}} & 43 (100\%) & 43 (100\%) & 7 / 11 \\
\addlinespace
Polyominoes $\uparrow$ & 8 (4 produced records) & +11.23 & 11.23 (100\%) & 11.23 (100\%) & 1 / 1 \\
\addlinespace
Signal Proc.\ $\uparrow$ & 6 (4 produced records) & +0.0186 & 0.0000 (0\%) & 0.0186 (100\%) & 3 / 4 \\
\addlinespace
Circle-Pack.\ $\uparrow$ & 4 (1 produced records) & +0.000093 & 0.000000 (0\%) & 0.000093 (100\%) & 2 / 4 \\
\addlinespace
Erd\H{o}s Over.\ $\downarrow$ & 6 (4 produced records) & \makecell{+0.001166 \\ {\scriptsize (C5 lower by 0.001166)}} & 0.000931 (80\%) & 0.001166 (100\%) & 1 / 2 \\
\addlinespace
EPLB $\uparrow$ & 6 (0 produced records) & +0.000154 & 0.000000 (0\%) & 0.000000 (0\%) & 0 / 2 \\
\addlinespace
LLM-SQL $\uparrow$ & 6 (2 produced records) & $-$0.001259 & --- & --- & 2 / 2 \\
\addlinespace
Txn Sched.\ $\uparrow$ & 3 (1 produced records) & +199.40 & 199.40 (100\%) & 199.40 (100\%) & 1 / 2 \\
\addlinespace
Cloudcast $\downarrow$ & 1 (0 produced records) & 0.0 (tie) & --- & --- & 0 / 2 \\
\bottomrule
\end{tabular}
\end{adjustbox}
\end{table}

Five phenomena follow from \Cref{tab:phenomena}. \textbf{(a)} Discovery leads the 4-agent baseline on 7 tasks. \textbf{(b)} On 6 of them (all but EPLB), the winning lineage contains directions the baseline never entered, and the entire gain comes from records built on them. On Signal Proc., Circle-Pack.\ and Erd\H{o}s Over.\ the own-record share is 0\%, 0\% and 80\%. There the final record falls in a shared direction but descends from Discovery-only ancestors. \textbf{(c)} On Kernel Eng., Polyominoes and Txn Sched.\ every record better than the baseline's final best falls in a Discovery-only direction (both columns 100\%). \textbf{(d)} The EPLB lead involves no new direction. All three methods lie within 0.1453--0.1455, at noise level. \textbf{(e)} LLM-SQL shows the cost of exploration. The directions Discovery opened peaked at 0.732704, below the baseline's locked-in direction (0.733963).

\section{Analysis: Two Why-Questions}\label{sec:analysis}

\Cref{sec:phenomena} left two questions, taken in turn here.

\subsection{Why does Discovery open directions the baseline never does?}\label{sec:why-open}

\begin{figure}[t]
  \centering
  \includegraphics[width=0.7\linewidth]{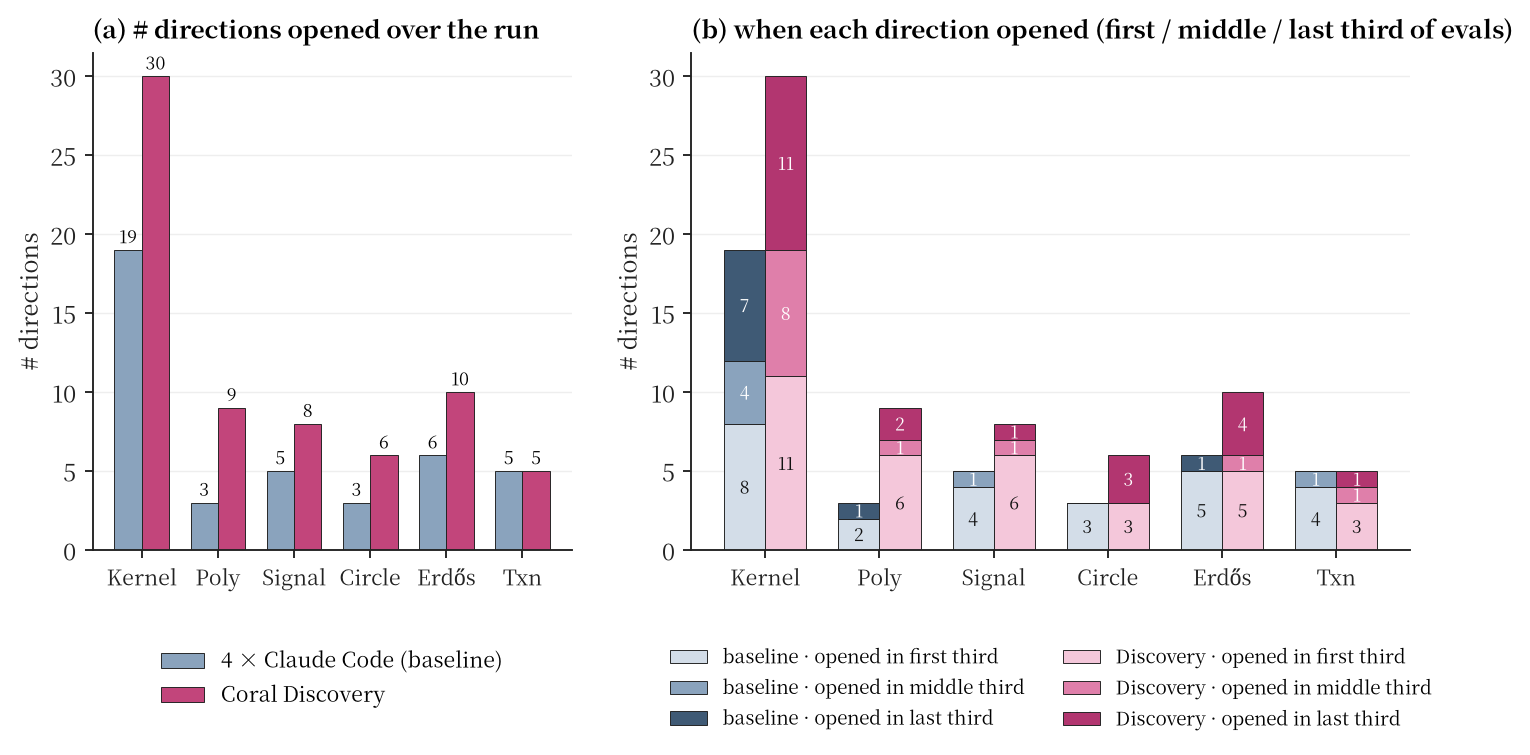}
  \caption{Directions opened on the six tasks in phenomena b and c (Kernel Eng., Polyominoes, Signal Proc., Circle-Pack., Erd\H{o}s Over., Txn Sched.), by re-labelled direction. (a) Over the whole run, and (b) in the first, middle and last third of the run. Blue-grey = 4-agent baseline, rose = \coral{}.}
  \label{fig:directions-opened}
\end{figure}

The direction board makes opening a direction part of the protocol. Agents open distinct directions in the first round. A direction holds at most two agents. The claim check rejects duplicate proposals. The NEW arm always prices one more direction, and the verifier registers new directions after every eval. Without these rules, the baseline's agents read the same leaderboard and crowd into one approach. \Cref{fig:directions-opened} counts the directions opened.

Discovery opened 68 directions to the baseline's 41, and it kept opening them: 34 against 15 after the first third (panel b). On all six tasks the gain rests on directions the baseline never entered (the ``lineage'' column in \Cref{sec:phenomena}). The 91.78 on Polyominoes comes from \dir{sa-shrink-boundary-removal}, opened only at eval 33.

\subsection{Why does Discovery's champion direction merge in more directions?}\label{sec:why-merge}

\begin{figure}[!t]
  \centering
  \begin{subfigure}[b]{0.4\linewidth}
    \centering
    \includegraphics[width=\linewidth]{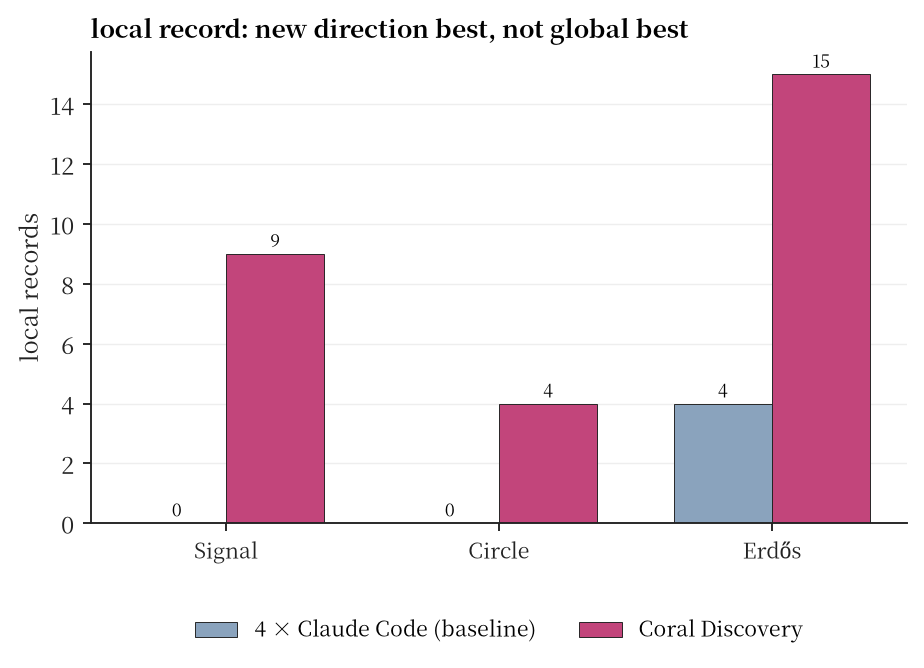}
    \caption{Local records (direction bests short of the global best) on the three tasks of phenomenon b.}
    \label{fig:local-records}
  \end{subfigure}\par\vspace{0.2cm}
  \begin{subfigure}[b]{1.0\linewidth}
    \centering
    \includegraphics[width=\linewidth]{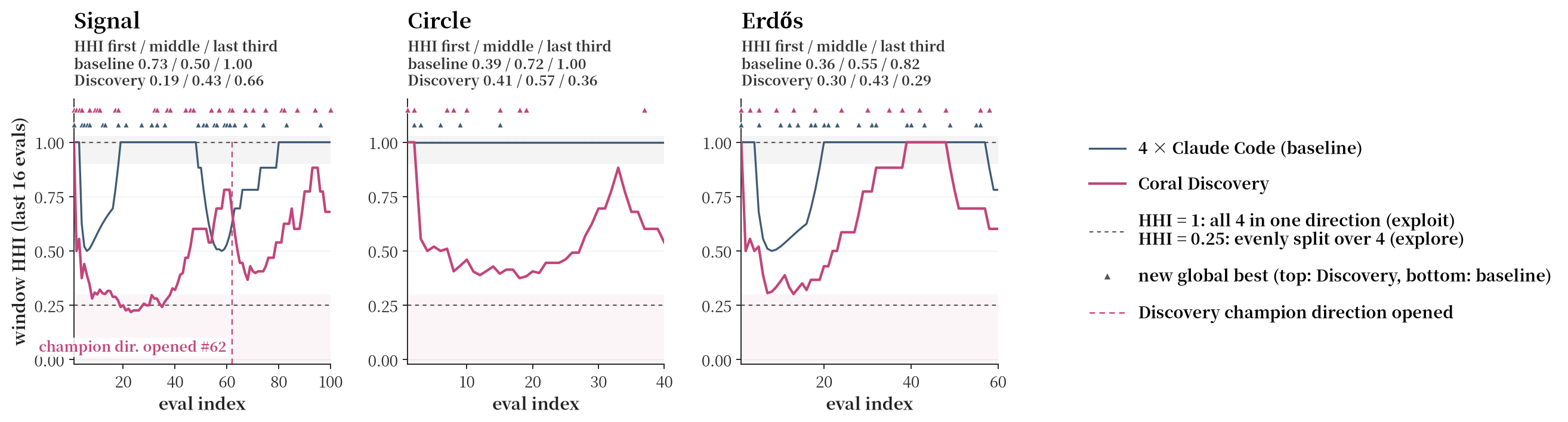}
    \caption{Window HHI over evals (window 16) on the same three tasks. Dashed lines mark HHI $=1$ (all four agents in one direction) and $0.25$ (even split), triangles mark new global bests (upper row Discovery, lower row baseline), and the rose vertical line marks the opening of Discovery's champion direction.}
    \label{fig:hhi}
  \end{subfigure}
  \caption{Lagging directions survive on the board and exploration continues alongside exploitation. Blue-grey = 4-agent baseline, rose = \coral{}.}
\end{figure}

On the three tasks of phenomenon b, Discovery's champion merged in other directions' results because a lagging direction does not die on the board. A node's credit is its improvement over the best of where it stands, $\max\bigl(0,\ \text{score} - \max(\text{parent's best}, \text{inspired-by node's best}, \text{own direction's best})\bigr)$, divided by the global best. A direction that progresses on its own track therefore keeps its yield rising and keeps its agents. Its node cards let others take its components. The baseline rewards only the global best and abandons whatever falls behind. Lagging directions on the board keep setting local records (direction bests short of the global best). Discovery vs baseline: Signal Proc.\ 9 vs 0, Circle-Pack.\ 4 vs 0 and Erd\H{o}s Over.\ 15 vs 4 (\Cref{fig:local-records}).

Exploration and exploitation proceed simultaneously (\Cref{fig:hhi}, window HHI over the last 16 evals). The baseline locks in and Discovery does not. HHI over the last third, baseline vs Discovery, is 1.00 vs 0.66 on Signal Proc., 1.00 vs 0.36 on Circle-Pack.\ and 0.82 vs 0.29 on Erd\H{o}s Over. All three champion directions were opened in the second half of the run (eval 62 / 100, 34 / 40 and 58 / 60). After the Signal Proc.\ champion's first global record, 76\% of the remaining 38 evals went to it. The HHI therefore rises but rarely reaches 1. The other two first records came fewer than 8 evals before the end.

\section{Limitations: Exploration Has a Cost}\label{sec:limitations}

Spreading the budget across directions is the source of every advantage above, and it has a price. On two tasks the price was not repaid. \coral{} finished below the 4-agent baseline on LLM-SQL and below the single agent on Txn Sched. \Cref{tab:lost} shows where each method spent its evals on these two tasks.

\begin{table}[!htbp]
\centering
\footnotesize
\setlength{\tabcolsep}{4pt}
\caption{Budget allocation on the two tasks Discovery did not lead. The fifth column is the share of evals in directions that never produced a global record, and the champion direction holds the final best.}
\label{tab:lost}
\begin{adjustbox}{max width=\linewidth}
\begin{tabular}{@{}llrrrll@{}}
\toprule
Task & Method & \makecell{Evals\\counted} & Directions & \makecell{Evals in directions that\\never produced a record} & Champion direction & Best @ eval \\
\midrule
LLM-SQL & Single Claude Code & 90 & 12 & 12\% & \makecell[l]{\dir{backward-tail-reopt} \\ {\scriptsize opened at eval 45, 43 evals, 8 records}} & 0.735769 {\scriptsize @81} \\
 & 4 $\times$ Claude Code baseline & 100 & 2 & 1\% & \makecell[l]{\dir{lexsort-greedy-prefix} \\ {\scriptsize opened at eval 1, 99 evals, 17 records}} & 0.733963 {\scriptsize @91} \\
 & \coral{} & 100 & 7 & 15\% & \makecell[l]{\dir{tree-row-greedy-col} \\ {\scriptsize opened at eval 12, 72 evals, 18 records}} & 0.732704 {\scriptsize @89} \\
\midrule
Txn Sched. & Single Claude Code & 100 & 2 & 0\% & \makecell[l]{\dir{simulated-annealing} \\ {\scriptsize opened at eval 2, 99 evals, 8 records}} & 4608 {\scriptsize @33} \\
 & 4 $\times$ Claude Code baseline & 100 & 5 & 0\% & \makecell[l]{\dir{neh-iterated-greedy} \\ {\scriptsize opened at eval 38, 59 evals, 6 records}} & 4367 {\scriptsize @98} \\
 & \coral{} & 100 & 5 & 18\% & \makecell[l]{\dir{adaptive-large-neighborhood-search} \\ {\scriptsize opened at eval 16, 23 evals, 4 records}} & 4566 {\scriptsize @71} \\
\bottomrule
\end{tabular}
\end{adjustbox}
\end{table}

\paragraph{LLM-SQL: opening directions brings headroom, but not necessarily the direction with headroom.}
\coral{} opened 7 directions where the baseline stayed in 2, and 15\% of its evals went to directions that never produced a record. Its champion, \dir{tree-row-greedy-col}, was opened at eval 12 and absorbed 72 evals for 18 records. Its best still stayed below the baseline's \dir{lexsort-greedy-prefix}, which produced 17 records in the same direction over 99 evals. The single agent did better than both. It opened \dir{backward-tail-reopt} at eval 45, a direction \coral{} never found. The per-direction credit rule that keeps lagging directions alive (\Cref{sec:why-merge}) is the same rule that kept agents in a direction with little headroom. Opening more directions raises the chance of finding headroom. Within a fixed budget, it does not guarantee that the direction with headroom is among those opened.

\paragraph{Txn Sched.: when one direction decides the outcome, a single agent holds the whole line.}
All three methods ended in a single champion direction, and the single agent won inside the direction it shared with the others. Its \dir{simulated-annealing}, opened at eval 2 and worked for 99 evals, reached 4608 at eval 33. \coral{} beat the 4-agent baseline through a direction the baseline never had, \dir{adaptive-large-neighborhood-search}. Its 4566 at eval 71 remained below the single agent. We can offer a conjecture. When the result is decided by method work inside one direction, a single agent carries the complete history of that direction. Every step was its own, and it knows which changes paid off. In a multi-agent system each agent inherits a direction through other agents' insights, cards and diffs. It receives the code but not the process. The four agents each walk only a short segment of the line.
\section{Conclusion}\label{sec:conclusion}

When several equal agents conduct autonomous research on the same problem, the gain in speed is immediate. The gain in diversity must be secured by how the community divides its cognitive labour. A shared leaderboard alone makes every agent a follower of the leader. \coral{} provides this structure with three components (\Cref{sec:method}). A verifier files every attempt under a direction. A direction board estimates the return of each direction and treats opening another direction as an option. A round protocol lets agents finish their own line first and then take the board as advice. On the nine \mitcoral{} tasks, with no internet access and without reading the grader, it outperforms the reproduced \mitcoral{} system on 6 tasks (\Cref{sec:results}). Under the same model and the same four-agent budget, it improves the Kernel Eng.\ result reported by \mitcoral{} from 1103 to 1095 cycles. It passes 1103 at eval 275, fewer than half of the 596 evals \mitcoral{} used. It also sets a new state of the art on Polyominoes (91.78, previously 87.0).

Labelling the direction of every submission shows that the leading scores come from directions the baseline never entered. These directions could be opened and survive until they produced records because of the combined effect of the claim rule, the NEW arm and per-direction credit (\Cref{sec:analysis}). The cost is equally clear (\Cref{sec:limitations}). A budget that is spread out does not necessarily land in a direction with headroom (LLM-SQL). When the outcome is decided by polishing the method inside a single direction, several agents taking over each other's process do not necessarily beat a single agent that walks the whole line itself (Txn Sched.).

\clearpage
\newpage
\bibliographystyle{plainnat}
\bibliography{references}

\clearpage
\appendix
\section{Command Set}\label{sec:commands}

Agents read and write shared state only through \texttt{coral} commands, since native \texttt{git} is disabled in their workspaces. \Cref{tab:commands} lists the complete agent-facing command set of \coral{} and its relation to \mitcoral{}. Five commands are new. \texttt{coral board} prints every direction's statistics, the NEW row, every agent's claim and the global best. \texttt{coral direction} lists one direction's attempts with their credit and insights. \texttt{coral claim} takes a seat on a direction, or opens a provisional one with \texttt{--new}, under the three checks of \Cref{sec:protocol}. \texttt{coral insight} is the only write of phase 2. \texttt{coral tree show} prints a node card. \texttt{coral eval} is extended with \texttt{--direction} (the declared direction, defaulting to the current claim) and \texttt{--inspired-by} (nodes whose code was borrowed, which enter the credit threshold). \texttt{coral checkout}, \texttt{show}, \texttt{diff} and \texttt{skills} are inherited unchanged. Four \mitcoral{} commands are withdrawn in this mode because the protocol replaces them. The leaderboard (\texttt{log}) is replaced by \texttt{board}, \texttt{direction} and \texttt{tree show}. Shared notes are replaced by the method and insight on every node card. \texttt{revert} is replaced by \texttt{checkout}. The heartbeat prompts are replaced by the Diagnose block (pivot) and the insight (reflect), with no counterpart for consolidation.

\begin{table}[h]
  \centering
  \caption{Complete agent-facing command set of \coral{} relative to \mitcoral{}. Five commands are new, \texttt{coral eval} is extended, four are inherited unchanged and four are withdrawn in this mode. Operator commands (\texttt{coral start}, \texttt{resume}, \texttt{stop}, \texttt{status}, \texttt{ui}) are unchanged and not used by agents.}
  \label{tab:commands}
  \small
  \begin{adjustbox}{max width=\linewidth}
  \begin{tabular}{@{}llp{0.40\linewidth}@{}}
    \toprule
    Status & Command & Description \\
    \midrule
    New & \texttt{coral board} & Global view: each direction's yield / potential / wanted / claimed, number of evals, best and top 3 attempts, the NEW row, each agent's claim, and the global best \\
    \addlinespace
    New & \texttt{coral direction \textless{}id\textgreater{} [--insights]} & All attempts of a single direction: score, credit, agent, parent, inspired-by, method, optionally with insights \\
    \addlinespace
    New & \makecell[tl]{\texttt{coral claim \textless{}id\textgreater{} --method "\ldots"} \\ \texttt{coral claim --new "\textless{}name\textgreater{}"} \\ \texttt{\quad --description "\ldots" --method "\ldots"}} & Takes a seat, subject to the three checks of \Cref{sec:protocol}. \texttt{--new} opens a provisional direction. Prints the other agents' current (direction, method) pairs \\
    \addlinespace
    New & \texttt{coral insight \textless{}commit\textgreater{} "\ldots"} & The only write operation of phase 2: writes an insight to the node card \\
    \addlinespace
    New & \texttt{coral tree show \textless{}commit\textgreater{}} & Node card: score, direction (declared / verdict), method, parent, inspired-by, credit and threshold, insight, children \\
    \addlinespace
    Extended & \makecell[tl]{\texttt{coral eval -m "\ldots" --direction \textless{}id\textgreater{}} \\ \texttt{\quad [--inspired-by c1,c2]}} & \mitcoral{}'s eval with two added parameters: the declared direction (defaults to the current claim) and the borrowed nodes \\
    \addlinespace
    Inherited & \makecell[tl]{\texttt{coral checkout \textless{}hash\textgreater{}} \\ \texttt{coral show \textless{}hash\textgreater{} [--diff]} \\ \texttt{coral diff} \\ \texttt{coral skills}} & Same as \mitcoral{}: reset the workspace to any attempt, view an attempt and its diff, view uncommitted changes, read shared skills \\
    \addlinespace
    Withdrawn & \makecell[tl]{\texttt{coral log} \\ \texttt{coral notes} \\ \texttt{coral revert} \\ \texttt{coral heartbeat}} & The leaderboard is replaced by \texttt{board}, \texttt{direction} and \texttt{tree show}. Notes are replaced by the method and insight on the node card. Revert is replaced by checkout. Heartbeat's pivot is replaced by the Diagnose block and its reflect by insight, and its consolidate has no counterpart \\
    \bottomrule
  \end{tabular}
  \end{adjustbox}
\end{table}

\end{document}